\documentclass{aa}
\usepackage{natbib}
\usepackage{graphicx}
\usepackage{txfonts}
\begin{document}
   \title{Estimation of absorption line indices of early-type galaxies using colours}


   \author{Z.~M. Li
          \inst{1, 2}
          \and
          R.~H. Li\inst{2}
          \and
          R.~X. Li\inst{2}}

   \offprints{Zhongmu Li}

   \institute{National Astronomical Observatories, Chinese
Academy of Sciences, Beijing 100012, China\\
              \email{zhongmu.li@gmail.com}
         \and
            Institute for Astronomy and History of Science and Technology, Dali University, Dali 671003, China\\
             }
   \titlerunning {Estimation of absorption line indices from colours of galaxies}
   \date{Received September 15, 2009; accepted March 16, 2009}


  \abstract
   {Absorption line indices are widely used to determine the stellar population parameters such as age and metallicity of galaxies,
   but it is not easy to obtain the line indices of some distant galaxies that have colours available. }
   {This paper investigates the correlations between absorption line indices and colours.}
   {A few statistical fitting methods are mainly used, via both the observational data of Sloan Digital Sky Survey and a widely used theoretical stellar population model.}
   {Some correlations between widely used absorption line indices and $ugriz$ colours are found from both observational data of early-type galaxies and a theoretical simple stellar population model.
   In particular, good correlations between colours and widely used absorption line indices such as D$_{\rm n}$(4000),
   H$\gamma_{\rm A}$, H$\gamma_{\rm F}$, H$\delta_{\rm A}$, Mg$_{\rm 1}$, Mg$_{\rm 2}$, and Mg$_{\rm b}$, are shown in this paper.}
   {Some important absorption line indices of early-type galaxies can be estimated from their colours
     using correlations between absorption line indices and colours. For example, age-sensitive absorption line indices can be
     estimated from $(u-r)$ or $(g-r)$ colours  and metallicity-sensitive
     ones from $(u-z)$ or $(g-z)$.
     This is useful for studying the stellar populations of distant galaxies, especially for statistical investigations.}


   \maketitle
%

\section{Introduction}

    The determination of stellar population properties (e.g., age, metallicity, and element abundances) of galaxies has long been an important
    subject in galaxy studies \citep{Tantalo:2004}. The evolutionary population synthesis technique has been widely used in such works \citep [see, e.g., ][]{Renzini:2006}.
    Many works confirm that some spectral features, which are called absorption line indices (the most famous ones are the well-known Lick indices), can
    reliably
    estimate stellar population properties \cite [e.g., ][]{Worthey:1994, Kong:2001, Bruzual:2003, Thomas:2003, Gallazzi:2005}.
    The definitions of widely used line strength indices can be seen in the papers of \citet {Bruzual:1983}, \citet{Worthey:1994}, \citet{Worthey:1997}, \citet{Huchra:1996},
    \citet{Diaz:1989}, \citet{Balogh:1999}, and
    \citet{Maraston:2009} among others. Although absorption line indices can determine the stellar
    population properties of galaxies well,
    this method cannot be used to study very distant (e.g., $z > 0.3$) galaxies because of the difficulty obtaining reliable spectral line indices.
    Meanwhile, some colours of such distant galaxies can be measured well. If some estimations of absorption line indices of galaxies can be
    derived from their colours, it will be able to investigate the stellar population properties of galaxies better, and then the
    formation and evolution of galaxies.

    This work proposes to study the correlations between colours and absorption line indices, and then
    presents a new method for estimating absorption line indices from the colours of galaxies. The organization of the
    paper is as follows. In Sect. 2, we briefly introduce the observational data and theoretical stellar population model used in this work. In Sect. 3,
    we study the correlations between colours and absorption line
    indices using the data of some early-type galaxies and theoretical stellar populations. Then the estimation of absorption line indices of galaxies from colours are discussed.
    Finally, we give our conclusions in Sect. 4.


\section{Observational data and theoretical stellar population model used in this work}

    This work uses the photometric data of Sloan Digital Sky Survey (SDSS, \citealt{SDSS:DR7})
    and a catalogue of absorption line indices of SDSS galaxies, which was built
    by the
    Max Plank Institute for Astrophysics (MPA) (see, e.g., \citealt{Gallazzi:2005}).
    SDSS is an imaging and spectroscopic survey of the high Galactic latitude
    sky visible from the northern hemisphere. The results of SDSS includes
    the photometry and spectroscopy of a large sample of galaxies. These data
    are reliable and have been widely used in studies in recent years.
    In addition, the catalogue of absorption line strength indices built by the MPA
    includes stellar absorption line indices corrected for sky lines
    using the best-fit model spectrum to interpolate over the sky line
    contamination. The catalogue supplies some stellar
    absorption line indices defined by \cite{Bruzual:1983}, \cite{Worthey:1994}, \cite{Worthey:1997}, \cite{Huchra:1996},
    \cite{Diaz:1989}, and \cite{Balogh:1999} for some of the galaxies observed by SDSS.
    Some of these indices are sensitive to stellar age (e.g., H$\beta$ and  D$_{\rm n}$(4000)), metallicity (e.g., Mg$_{\rm b}$), or element abundances, and are therefore usually used to
    determine the stellar population properties of galaxies. One
    can read \cite{Gallazzi:2005} as an example. Because the
    catalogue supplies the absorption line indices of galaxies and the photometric data of
    these galaxies are available in the SDSS, we chose
    the catalogue as a data source of this study.

    The absorption line indices and photometric data of each galaxy is
    combined according to some special identifiers (objID and SpecobjID) of galaxies. One
    can refer to the web pages of the SDSS for more details.
    When selecting sample galaxies, only galaxies with small uncertainties
    in magnitudes and absorption line indices are chosen in our working
    samples. This guarantees that observational uncertainties in colours and absorption line indices
    lead to a small uncertainty in the final results.
    All sample galaxies have magnitude uncertainties less than 0.05\,mag, and the uncertainty of each absorption
    line index is less than the average uncertainty of the line indices of all galaxies in the Fourth Data Release of SDSS
    (SDSS-DR4). As a result, 2287 early-type galaxies with concentration index $C \geq$ 2.6 were chosen
    for our working galaxy sample. When calculating the colours of galaxies, the values for $k$-corrections and galactic extinctions supplied by
    SDSS database were used directly.

    We chose the widely used simple stellar population model of \cite{Bruzual:2003} (hereafter BC03 model) for this
    work. BC03 model is a theoretical stellar population model that has been compared to the main features of galaxies observed by
    SDSS,
    and it has been widely used in all kinds of stellar population studies. The model supplies both $ugriz$
    colours and absorption line indices of simple stellar populations under a spectral resolution close to SDSS.
    Thus it is convenient to use the model with the data of SDSS for this investigation.

\section{Correlations between colours and absorption line indices}

\subsection{Observational results}
    Using the data of $ugriz$ colours and absorption line indices of 2287 early-type galaxies,
    we investigated the correlations between 26 widely used absorption line indices and colours.
    The results show that most absorption line indices correlate to
    some colours. We therefore tried to find the best colour for estimating each line index
    by comparing the correlations of all pairs consisting of a line index and a colour (hereafter line index-colour pair).
    Because 26 absorption line indices and 10 colours respecting
    $u$, $g$, $r$, $i$, and $z$ magnitudes are taken into account,
    the correlations for 260 line index-colour pairs are investigated in this work.
    Then for each absorption line index, we find the
    correlation with the least scatter from all correlations relating to
    it. The results show that in 26 line
    indices, 20 correlate well with some colours.
    In Figs. 1--5, the best-fit correlations are compared with observational
    data. The detailed best-fit correlations are then listed in
    Table 1, together with their 1 $\sigma$ scatters. Because the data of a
    few (about 15 with blue colours) galaxies do not appear statistically
    significative, they are not used for giving the final fitting correlations between colours and absorption line
    indices. This results in the various galaxy samples used for
    studying each pair of line index and colour.
    In order to give results,
    any correlation relating to one line index that has the least scatters when using the whole sample (2287 galaxies) for fitting
    is then taken as the best-fit correlation  of the line index (see Table
1), although this correlation shows bigger fitting uncertainties
    when taking smaller (excluding galaxies with significant data) galaxy samples.

   \begin{figure}
   \centering
   \includegraphics[angle=-90,width=88mm]{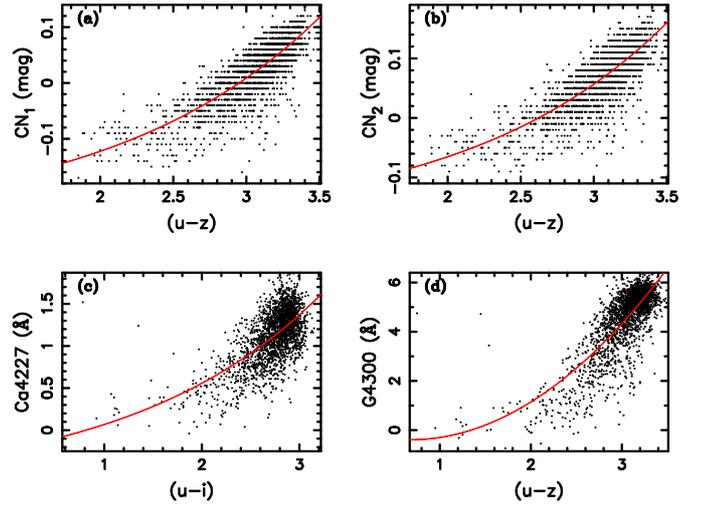}
      \caption{Comparison of original (circles) and best-fit (lines)
      correlations between four absorption line indices and colours of a sample of 2287 early type galaxies in SDSS.
      Panels a), b), c), and d) are for CN$_{1}$, CN$_{2}$, Ca4227,
      and G4300, respectively.
              }
         \label{Fig1}
   \end{figure}
%

   \begin{figure}
   \centering
   \includegraphics[angle=-90,width=88mm]{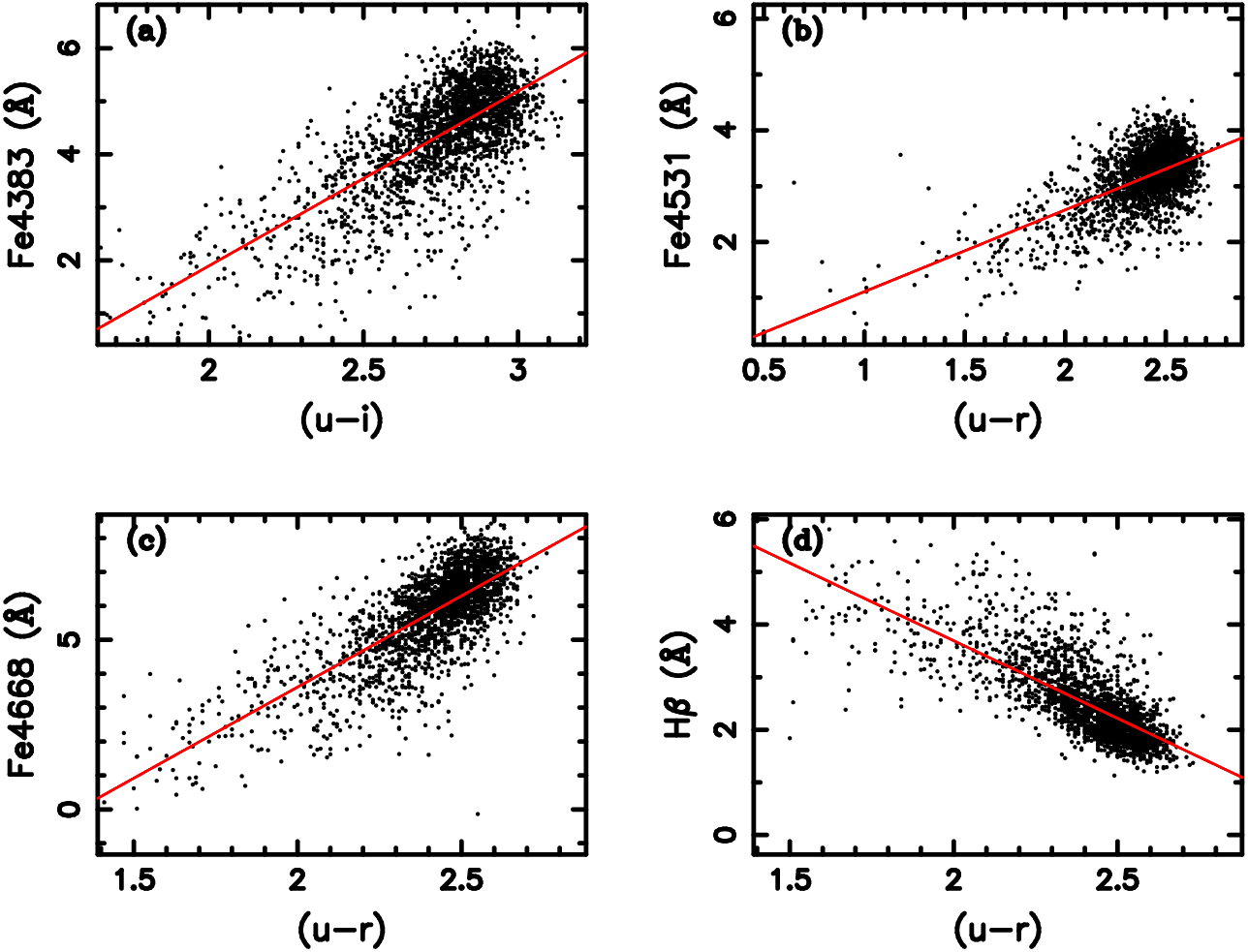}
      \caption{Similar to Fig. 1, but for absorption line indices Fe4383, Fe4531,
      Fe4668, and H$\beta$.
              }
         \label{Fig2}
   \end{figure}
%

   \begin{figure}
   \centering
   \includegraphics[angle=-90,width=88mm]{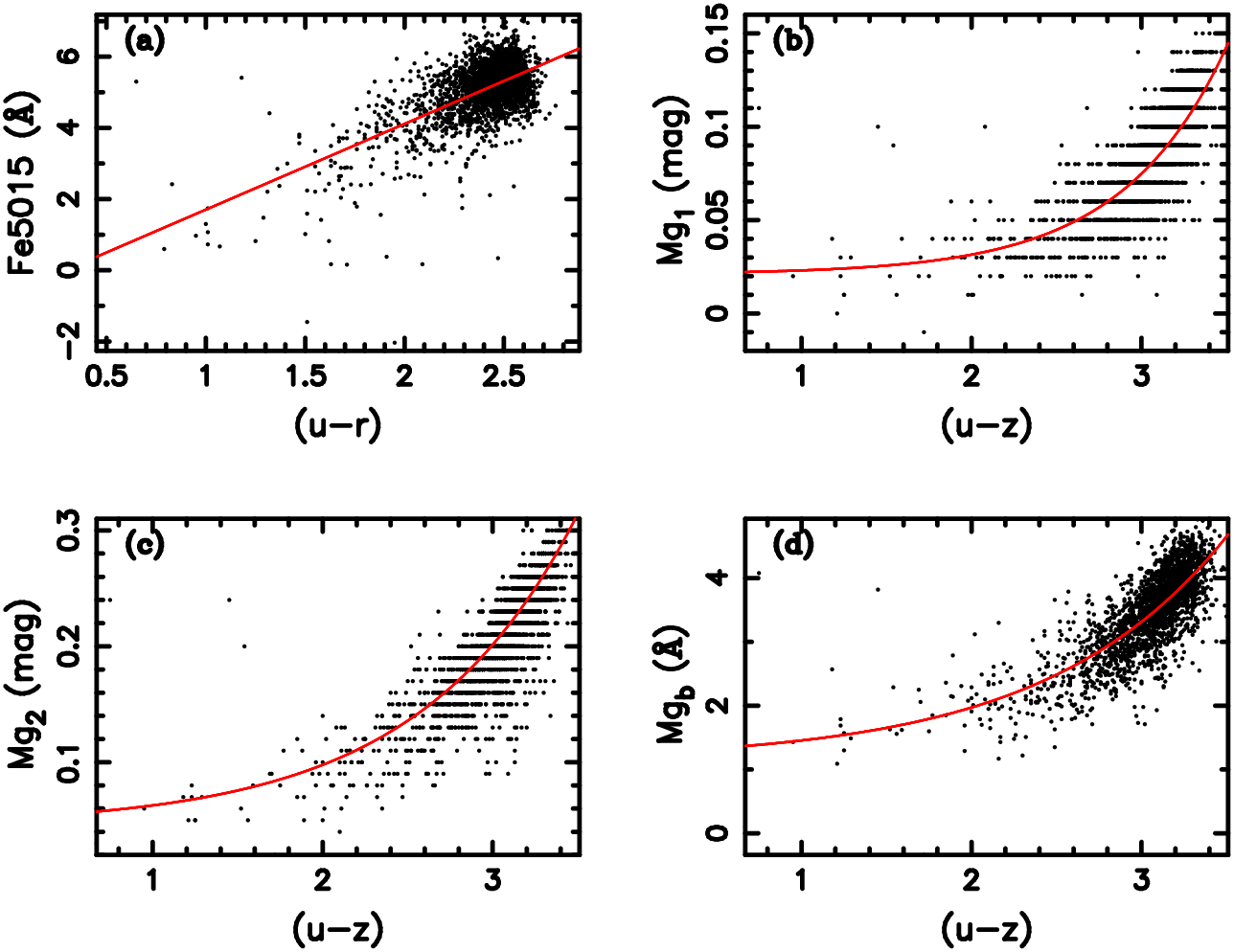}
      \caption{Similar to Fig. 1, but for absorption line indices Fe5015,
      Mg$_{\rm 1}$, Mg$_{\rm 2}$, and Mg$_{\rm b}$.
              }
         \label{Fig3}
   \end{figure}
%

   \begin{figure}
   \centering
   \includegraphics[angle=-90,width=88mm]{15952fg4.ps}
      \caption{Similar to Fig. 1, but for absorption line indices Fe5270,
      Fe5335, Na$_{\rm D}$, and H$\delta_{\rm A}$.
              }
         \label{Fig4}
   \end{figure}
%

   \begin{figure}
   \centering
   \includegraphics[angle=-90,width=88mm]{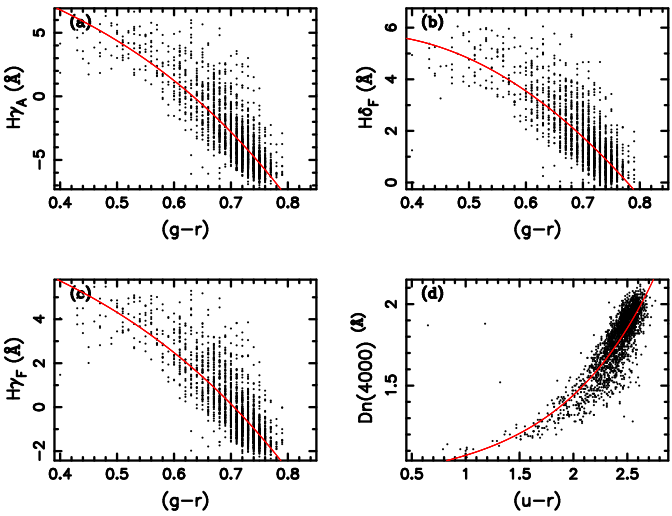}
      \caption{Similar to Fig. 1, but for absorption line indices H$\gamma_{\rm A}$,
      H$\delta_{\rm F}$, H$\gamma_{\rm F}$, and D$_{\rm n}$(4000).
              }
         \label{Fig5}
   \end{figure}
%

\begin{table}[]
\caption[]{Best fit correlations between 20 absorption line indices
and $ugriz$ colours of early-type galaxies in SDSS.} \label{Tab1}
\begin{center}\begin{tabular}{lccc}
\hline \hline\noalign{\smallskip}
Best fit Correlation  &$\sigma$ &Unit\\    
\hline
\\
CN$_{\rm 1}$=-0.254+0.033~e$^{(u-z)/1.446}$  &0.031  &mag\\
CN$_{\rm 2}$=-0.180+0.027~e$^{(u-z)/1.382}$  &0.030  &mag\\
Ca4227=-0.678+0.450~e$^{(u-i)/1.980}$  &0.242   &$\AA$\\
G4300=-0.001-1.170~$(u-z)$+0.871~$(u-z)^{2}$   &0.807   &$\AA$\\

Fe4383=-4.697+3.296~$(u-i)$  &0.707   &$\AA$\\
Fe4531=-0.357+1.466~$(u-r)$  &0.442   &$\AA$\\
Fe4668=-7.147+5.376~$(u-r)$  &0.904   &$\AA$\\
H$\beta$=9.596-2.952~$(u-r)$ &0.515   &$\AA$\\

Fe5015=-0.703+2.408~$(u-r)$   &0.696   &$\AA$\\
Mg$_{\rm 1}$=~0.021+0.0004~e$^{(u-z)/0.612}$  &0.016   &mag\\
Mg$_{\rm 2}$=~0.045+0.006~e$^{(u-z)/0.920}$  &0.026   &mag\\
Mg$_{\rm b}$=~1.133+0.124~e$^{(u-z)/1.046}$ &0.411   &$\AA$\\

Fe5270=0.021+0.855~$(u-z)$  &0.356   &$\AA$\\
Fe5335=0.468+0.636~$(u-z)$  &0.394   &$\AA$\\
Na$_{\rm D}$=1.379+0.061~e$^{(u-z)/0.934}$  &0.580   &$\AA$\\
H$\delta_{\rm A}$=13.874-1.906~e$^{(g-r)/0.359}$  &1.415   &$\AA$\\

H$\gamma_{\rm A}$=15.154-3.011~e$^{(g-r)/0.392}$  &1.548   &$\AA$\\
H$\delta_{\rm F}$=3.266+16.141~$(g-r)$-26.131~${(g-r)}^2$  &0.746   &$\AA$\\
H$\gamma_{\rm F}$=10.898-1.934~e$^{(g-r)/0.408}$  &0.921   &$\AA$\\
D$_{\rm n}$(4000)=0.892+0.059~e$^{(u-r)/0.896}$  &0.102   &$\AA$\\
\\
 \noalign{\smallskip}\hline
\end{tabular}\end{center}
\tablefoot{The results are derived from a sample of 2287 galaxies.
All colours are in mag. Three columns are for best-fit correlations
between absorption line indices and $ugriz$ colours, 1 $\sigma$
uncertainties (``$\sigma$'' in the table), and the units of line
indices.}
\end{table}

    As we see, Figs. 1--5 show that there are good correlations
    between most absorption line indices and colours. Compared to Table
    1, it is found that these correlations can be fitted well via linear,
    exponential, or polynomial correlations. However, as
    we see, most of the best-fit correlations relate to the $u$ band.
    According to the report of the SDSS, magnitudes in $u$ band usually
    have larger uncertainties than others. Thus it is
    better to take colours relating only to $griz$ bands
    when using SDSS data for
    research. We therefore find the best-fit correlations between
    absorption line indices and the colours only relating to $griz$
    magnitudes. The results are shown in Figs. 6-9 and Table 2.
    The results of H$\delta_{\rm A}$, H$\gamma_{\rm A}$,
    H$\delta_{\rm F}$ and H$\gamma_{\rm F}$ are not repeated here,
    because one can refer to Figs. 1--5 and Table 1 for details of their results.

   \begin{figure}
   \centering
   \includegraphics[angle=-90,width=88mm]{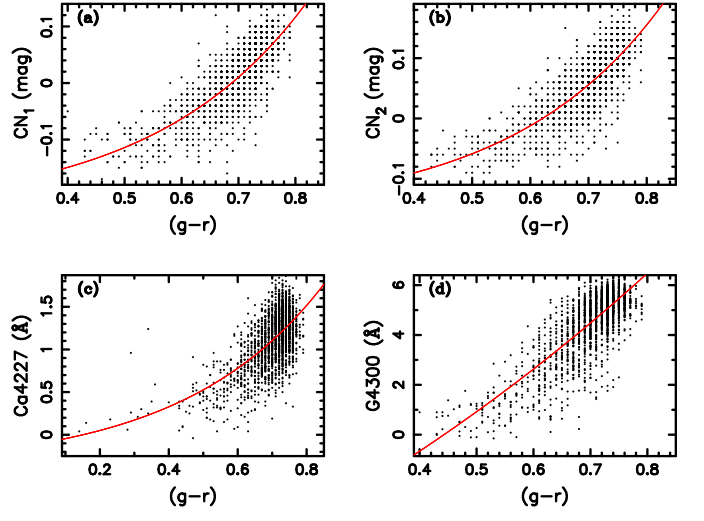}
      \caption{Similar to Fig. 1, but for correlations between $griz$ colours and absorption line indices.
      Panels a), b), c), and d) are for CN$_{1}$, CN$_{2}$, Ca4227,
      and G4300, respectively.
              }
         \label{Fig6}
   \end{figure}
%

   \begin{figure}
   \centering
   \includegraphics[angle=-90,width=88mm]{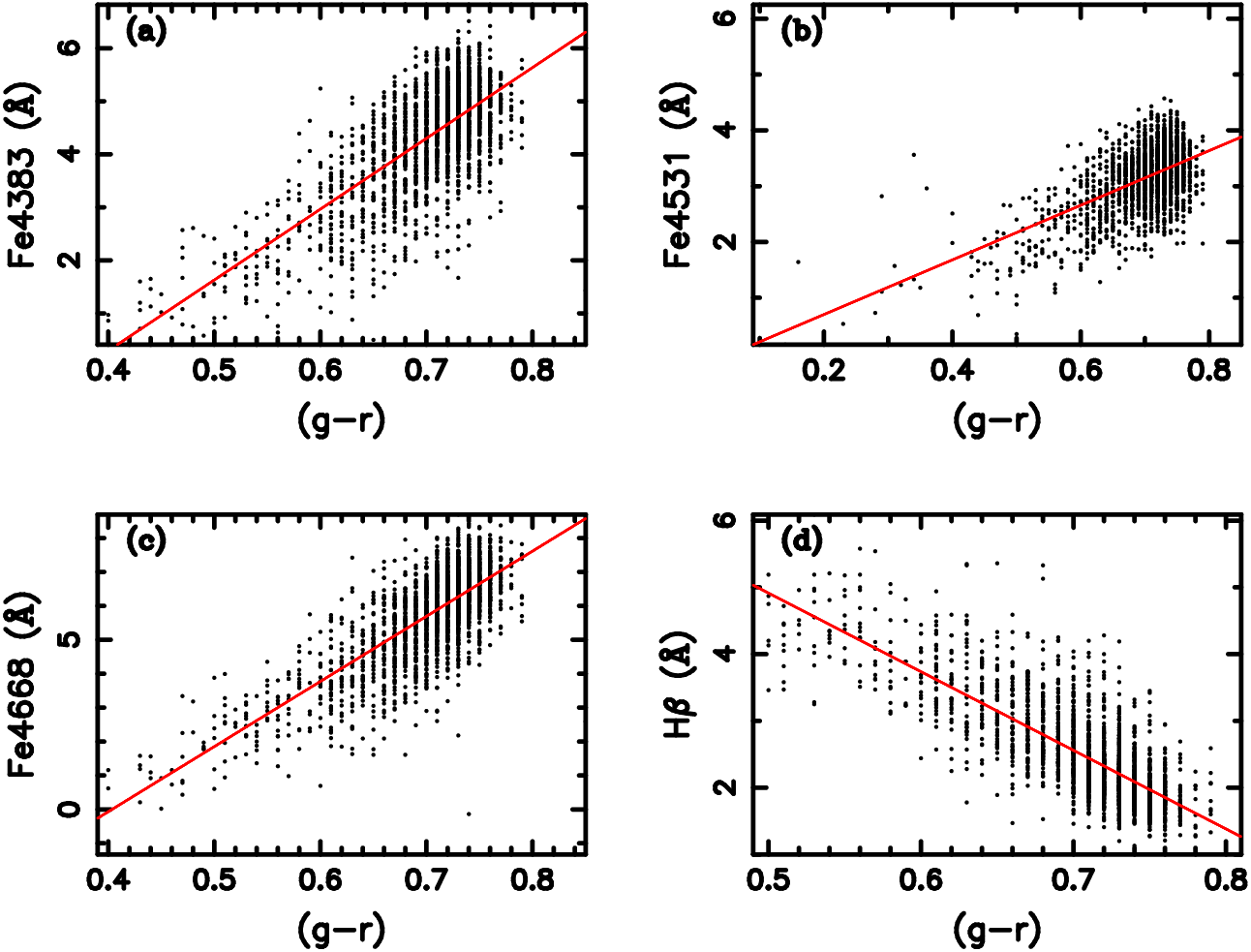}
      \caption{Similar to Fig. 6, but for absorption line indices Fe4383, Fe4531,
      Fe4668, and H$\beta$.
              }
         \label{Fig7}
   \end{figure}
%

   \begin{figure}
   \centering
   \includegraphics[angle=-90,width=88mm]{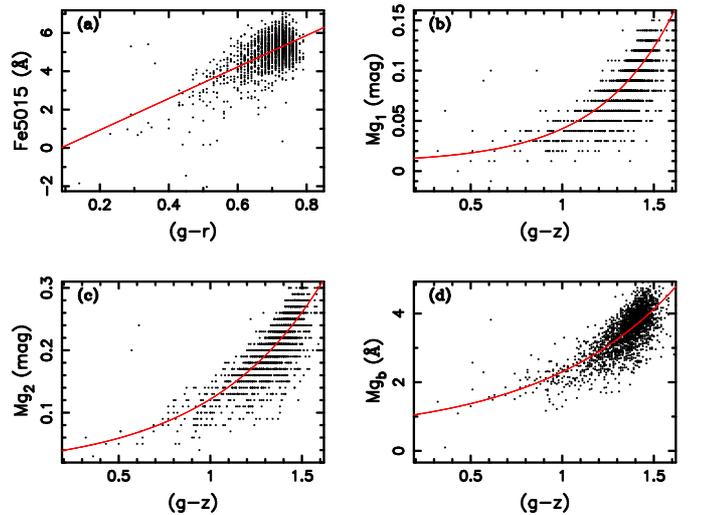}
      \caption{Similar to Fig. 6, but for absorption line indices Fe5015,
      Mg$_{\rm 1}$, Mg$_{\rm 2}$, and Mg$_{\rm b}$.
              }
         \label{Fig8}
   \end{figure}
%

   \begin{figure}
   \centering
   \includegraphics[angle=-90,width=88mm]{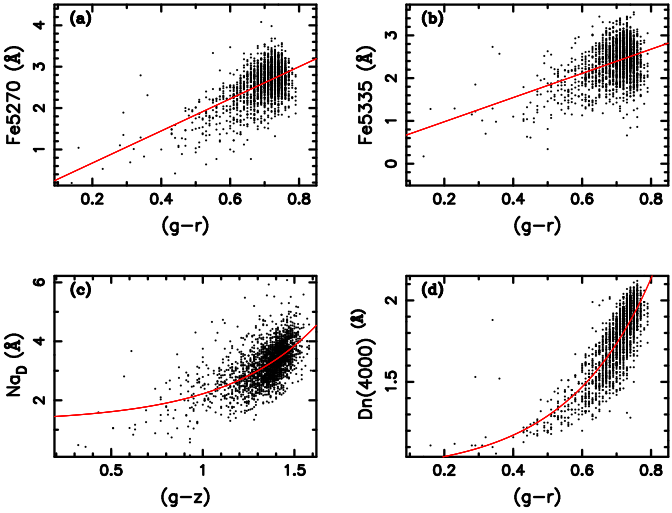}
      \caption{Similar to Fig. 6, but for absorption line indices Fe5270,
      Fe5335, Na$_{\rm D}$, and D$_{\rm n}$(4000).
              }
         \label{Fig9}
   \end{figure}
%

\begin{table}[]
\caption[]{Similar to Table 1, but for best-fit correlations between
16 absorption line indices and $griz$ colours.} \label{Tab2}
\begin{center}\begin{tabular}{lccc}
\hline \hline\noalign{\smallskip}
Correlation  &$\sigma$  &Unit\\    
\hline
\\
CN$_{\rm 1}$=-0.228+0.018~e$^{(g-r)/0.271}$  &0.032  &mag\\
CN$_{\rm 2}$=-0.159+0.015~e$^{(g-r)/0.263}$  &0.031  &mag\\
Ca4227=-0.391+0.271~e$^{(g-r)/0.410}$  &0.250  &$\AA$\\
G4300=-20.271+14.374~e$^{(g-r)/1.288}$   &0.784  &$\AA$\\
Fe4383=-5.047+13.353~$(g-r)$  &0.718  &$\AA$\\
Fe4531=-0.279+ 4.893~$(g-r)$  &0.454  &$\AA$\\
Fe4668=-7.753+19.206~$(g-r)$    &0.898  &$\AA$\\
H$\beta$=10.801-11.774~$(g-r)$  &0.466  &$\AA$\\
Fe5015=-0.715+8.233~$(g-r)$   &0.707  &$\AA$\\
Mg$_{\rm 1}$=~0.008+0.003~e$^{(g-z)/0.411}$  &0.018  &mag\\
Mg$_{\rm 2}$=~0.008+0.023~e$^{(g-z)/0.626}$  &0.028  &mag\\
Mg$_{\rm b}$=~0.420+0.486~e$^{(g-z)/0.738}$ &0.429  &$\AA$\\
Fe5270=-0.103+3.879~$(g-r)$  &0.359  &$\AA$\\
Fe5335=~0.415+2.828~$(g-r)$  &0.397  &$\AA$\\
Na$_{\rm D}$=1.253+0.134~e$^{(g-z)/0.506}$  &0.553  &$\AA$\\
D$_{\rm n}$(4000)=0.936+0.047~e$^{(g-r)/0.247}$  &0.114  &$\AA$\\
\\
 \noalign{\smallskip}\hline
\end{tabular}\end{center}
\tablefoot{The results for H$\delta_{\rm A}$, H$\gamma_{\rm A}$,
H$\delta_{\rm F}$, H$\gamma_{\rm F}$ are listed in Table 1. The
somewhat smaller fitting uncertainties of G4300, Fe4668, H$\beta$,
and Na${\rm _D}$ result from fewer galaxies (about 2270) used for
fitting compared to the results of Table 1.}
\end{table}

\subsection{Results from theoretical stellar population model}
   The previous section shows some good correlations between
   observational absorption line indices and $ugriz$ colours of early-type galaxies.
   However, the reliability of these
   correlations remains unclear. If these correlations can be found in theoretical
   stellar population models, they should be believable.
   This section tries to search for similar
   correlations from the theoretical stellar population model of \cite{Bruzual:2003}.
   Because early-type galaxies are close to simple stellar populations,
   we take the data of some simple stellar populations (with 6 metallicities and 220 ages) for our study.
   It shows that there are good correlations between
   some absorption line indices and $ugriz$ colours of the BC03 model.
   The main results can be seen in Figs. 10 to 12, in which the results
   of a few absorption line indices that correlate well to colours are shown.
   We see that there
   are clear correlations between $ugriz$ colours and CN$_{\rm 1}$, CN$_{\rm 2}$,
   Ca4227, G4300, Mg$_{\rm 1}$, Mg$_{\rm 2}$, Mg$_{\rm b}$, H$\delta_{\rm A}$,
   H$\gamma_{\rm A}$, H$\delta_{\rm F}$, H$\gamma_{\rm
   F}$, and D$_{\rm n}$(4000), for stellar populations within the colour ranges listed in Table 3.

   \begin{figure}
   \centering
   \includegraphics[angle=-90,width=88mm]{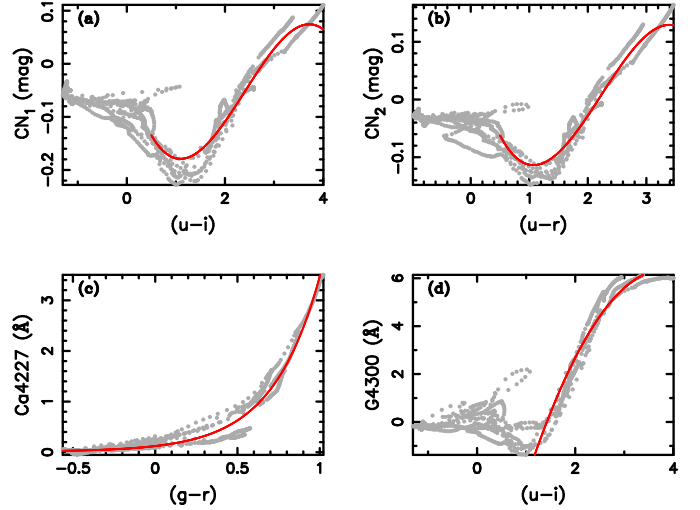}
      \caption{Correlations between absorption line indices and $ugriz$ colours of the theoretical stellar populations in the BC03 model \citep{Bruzual:2003}. Points in each absorption line index versus colour plane show stellar populations with 220 ages and 6 metallicities. Lines are fitted correlations.
      Panels from a) to d) are for  CN$_{\rm 1}$, CN$_{\rm 2}$, Ca4227, and G4300,
      respectively.
              }
         \label{Fig10}
   \end{figure}
%

   \begin{figure}
   \centering
   \includegraphics[angle=-90,width=88mm]{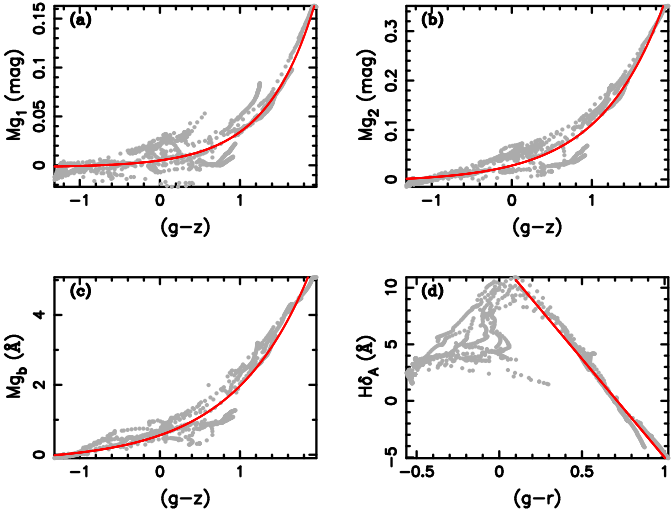}
      \caption{Similar to Fig. 10, but for Mg$_{\rm 1}$, Mg$_{\rm 2}$, Mg$_{\rm b}$, and H$\delta
    _{\rm A}$.
              }
         \label{Fig11}
   \end{figure}
%

   \begin{figure}
   \centering
   \includegraphics[angle=-90,width=88mm]{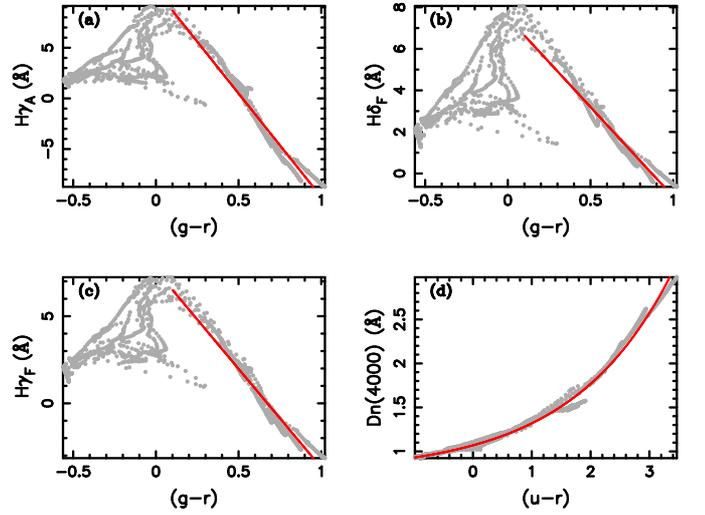}
      \caption{Similar to Fig. 10, but for  H$\gamma_{\rm A}$, H$\delta_{\rm F}$, H$\gamma_{\rm
   F}$, and D$_{\rm n}$(4000).
              }
         \label{Fig12}
   \end{figure}
%

\begin{table*}[]
\caption[]{Best-fit correlations between 12 absorption line indices
and the $ugirz$ colours of stellar populations of the BC03 model.}
\label{Tab3}
\begin{center}\begin{tabular}{lccc}
\hline \hline\noalign{\smallskip}
Best Fit Correlation  &$\sigma$   &Colour Range [mag]\\    
\hline
\\
CN$_{\rm 1}$=-0.011-0.341~$(u-i)$+0.202~$(u-i)^{2}$-0.028~$(u-i)^{3}$ &0.026 & ~0.50$\leq(u-i)\leq$4.00\\
CN$_{\rm 2}$=0.090-0.425~$(u-r)$+0.261~$(u-r)^{2}$-0.039~$(u-r)^{3}$ &0.020 & ~0.50$\leq(u-r)\leq$3.47\\
Ca4227=0.012+0.110~e$^{(g-r)/0.290}$  &0.140   & -0.56$\leq(g-r)\leq$1.02\\
G4300=-9.681+8.353~$(u-i)$-1.087~$(u-i)^{2}$ &0.584 & ~1.00$\leq(u-i)\leq$4.00\\
Mg$_{\rm 1}$=-0.002+0.007~e$^{(g-z)/0.610}$  &0.012 & -1.32$\leq(g-z)\leq$1.95\\
Mg$_{\rm 2}$=-0.005+0.033~e$^{(g-z)/0.795}$  &0.021 & -1.32$\leq(g-z)\leq$1.95\\
Mg$_{\rm b}$=-0.206+0.764~e$^{(g-z)/0.956}$  &0.298 & -1.32$\leq(g-z)\leq$1.95\\
H$\delta_{\rm A}$=12.436-17.446~$(g-r)$ &1.050 & ~0.10$\leq(g-r)\leq$1.02\\
H$\gamma_{\rm A}$=10.701-20.424~$(g-r)$ &1.214 & ~0.10$\leq(g-r)\leq$1.02\\
H$\delta_{\rm F}$=7.503-8.613~$(g-r)$   &0.642 & ~0.10$\leq(g-r)\leq$1.02\\
H$\gamma_{\rm F}$=7.639-11.328~$(g-r)$  &0.677 & ~0.10$\leq(g-r)\leq$1.02\\
D$_{\rm n}$(4000)=0.754+0.315~e$^{(u-r)/1.711}$  &0.051  & -0.99$\leq(u-r)\leq$3.47\\
\\
 \noalign{\smallskip}\hline
\end{tabular}\end{center}
\end{table*}

\subsection{Comparison of colour-absorption line index correlations derived from observational data and the theoretical model}
    Because both the observational data and the theoretical stellar
    population model show good correlations between
    absorption line indices and colours, we compare two kinds of results here.
    The detailed comparisons are shown in Figs. 13
    and 14.
    As can be seen, both observational data and
    the BC03 model show similar trends for how absorption
    line indices change with colours.
    In particular, both observational data and the BC03
    model show exponential correlations for three metallicity-sensitive absorption line indices (Mg$_{\rm 1}$,
    Mg$_{\rm 2}$, Mg$_{\rm b}$) and
    colours (Fig. 13), and the scatters of observational and theoretical correlations are very close.
    However, the correlations between four
    age-sensitive line indices and colours, which were derived from
    observational data and BC03 model, are shown to be different (Fig. 14).
    When the observational data show some exponential
    correlations, the data of BC03 model report some linear
    correlations for Balmer lines and colours. In addition, although both the observational data and BC03 model give
    exponential correlations for D$_{n}$(4000) and $(u-r)$, the two correlations show clear differences.
    The differences between the observational data of SDSS and theoretical stellar population model possibly
    result from the limitation of the stellar population model and observational uncertainties.
    When we tried to find the correlation between D$_{n}$(4000) and $(g-r)$ of BC03 stellar populations,
    the result showed that there is no clear correlation between the two
    indices;
    however, the observational data imply a correlation between them, as can be seen in Fig. 9.
    This suggests that simple stellar populations in the BC03 model are different from the sample galaxies.
    Therefore, the internal difference between simple stellar population models
    and real galaxies should mainly contribute to the differences between correlations derived from observational data and
    theoretical stellar populations.

    As a whole, the observational data of SDSS and the BC03 theoretical stellar population model give similar
    trends for the correlations between absorption line indices and colours. Thus both correlations
    derived from observational data and theoretical stellar population model can be used for future investigations,
    especially for some statistical investigations.

   \begin{figure}
   \centering
   \includegraphics[angle=-90,width=88mm]{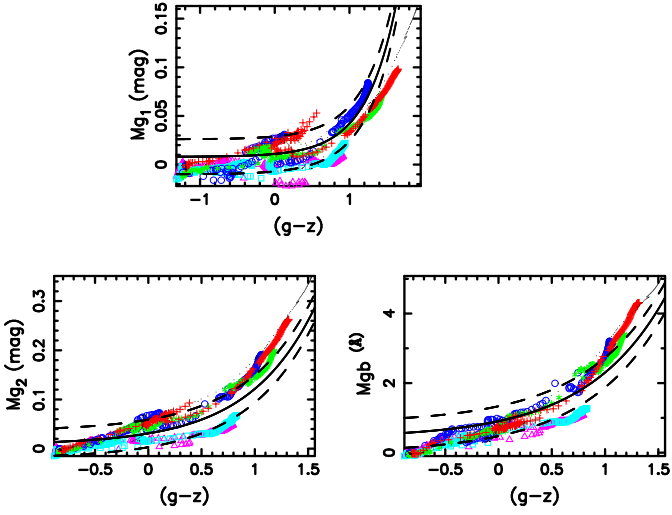}
      \caption{Comparison of correlations between absorption line indices and colours derived from the observational data and BC03 model, respectively. Points show the correlations derived from BC03 model.
      Six kinds of points denote stellar populations wit metallicities $Z$ = 0.0001,0.0004,0.004,0.008, 0.02, and 0.05.
      Solid lines show the best fits to correlations between observational absorption line indices and colours of early-type galaxies.
      Dashed lines show the 1 $\sigma$ scatters of correlations derived from observational data.
      Three panels are for three metallicity sensitive indices, i.e., Mg$_{\rm 1}$, Mg$_{\rm 2}$, and Mg$_{\rm b}$,
      respectively.
              }
         \label{Fig13}
   \end{figure}
%

   \begin{figure}
   \centering
   \includegraphics[angle=-90,width=88mm]{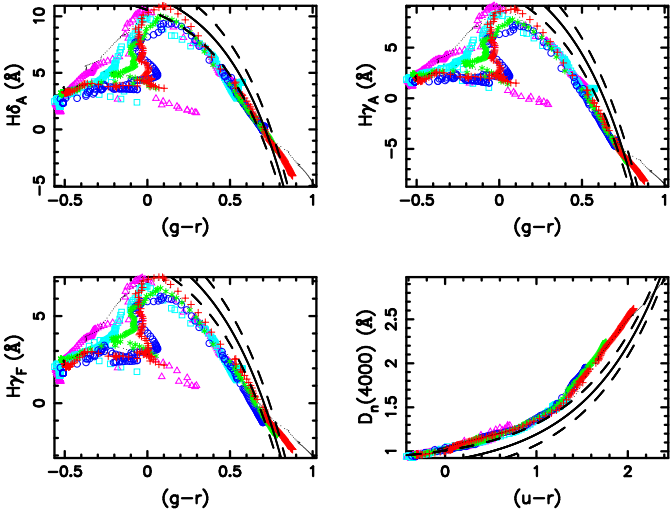}
      \caption{Similar to Fig. 13, but for four age-sensitive indices, H$\delta_{\rm A}$,
      H$\gamma_{\rm A}$, H$\gamma_{\rm F}$, and D$_{\rm n}$(4000).
              }
         \label{Fig14}
   \end{figure}
%

\subsection{Potential applications of results and discussions}

    It is shown that there are good
    correlations between most absorption line indices and colours of
    galaxies. Therefore, it is possible to measure absorption line indices of
    galaxies that have no reliable spectral data, but with colours available,
    and then to give stellar population studies to some distant galaxies.

    The correlations derived from the observational data of SDSS are actually more
    suitable for statistical or relative studies
    because of the somewhat large uncertainties of fitted correlations.
    In addition, the final correlations are more suitable for studying early-type galaxies,
    because our results are derived from an early-type galaxy sample.
    However, the results can also work for
    relative studies of all galaxies. The reason is
    that both early and late type galaxies obey some similar correlations of
    absorption line indices and colours. In Fig. 15, a few
    correlations derived from different (early, late, mixed) galaxy samples are compared.
    It is shown that, although there are differences among the
    correlations derived from three galaxy samples, they give
    similar trends for the change of line indices with colours.

    As expected, the correlations derived from the observational data and the BC03
    model makes it possible to measure some absorption line indices
    of galaxies from colours. When
    using the results, one needs to take the colour ranges (see Table 3) of stellar
    populations into account. In addition, the results are actually
    model dependent. Different results may be obtained using
    correlations derived from various stellar population models.

    Although absorption line indices correlate to colours, and
    it is possible to estimate line indices of galaxies,
    the application of these correlations depends on the reliability of colour determinations.
    Because $K$ corrections of the magnitudes can affect the magnitudes directly,
    it is very important to take the uncertainty of $K$ correction into account when using
    colours to estimate the absorption line indices of galaxies. This will determine what galaxies the correlations found in this
    paper can be used for studying. To make this clear, we investigated
    how the $K$ correction uncertainty changes with some of the properties of galaxies.
    In detail, the relations between uncertainty in $K$ correction and redshift,
    magnitude, concentration index, a/b ratio, $K$ correction,
    redshift error, and magnitude error are investigated.
    The uncertainty of $K$ correction is calculated from
    redshift uncertainty, according to the determination of the $K$ correction in the SDSS process.
    As a result, the uncertainty in $K$ correction is found to correlate to the a/b ratio,
    Petrosian magnitude in $r$ band, and magnitude error in $r$ band. However, there is no simple changing
    trend between $K$ correction uncertainty and redshift.
    More detailed results can be seen in Figs. 16 to 19. These
    results will be helpful for estimating the $K$ correction uncertainty of each galaxy,
    and then for the effects on determining absorption line indices. Because
    it shows that $K$ correction uncertainty is strongly correlated
    to magnitude error, the result suggests using magnitude error for
    estimating the $K$ correction uncertainties of galaxies.

    Furthermore, the 1 $\sigma$ uncertainties in fitting correlations are shown to be different for galaxy samples within various redshift ranges. Table 4 shows the detailed results. It will be useful for estimating the final uncertainties when the fitting correlations are used to calculate the absorption line indices of galaxies with various distances.

   \begin{figure}
   \centering
   \includegraphics[angle=-90,width=88mm]{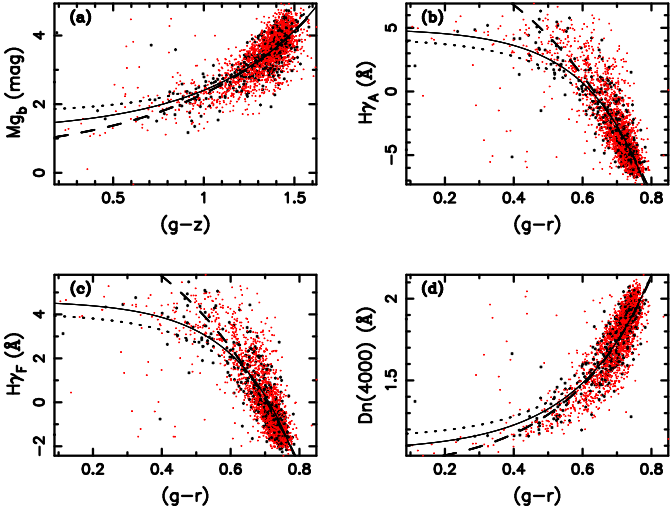}
   \caption{Comparison of correlations between absorption line indices and colours,
     which are derived from early-type galaxy sample (dashed),
     late-type galaxy sample (dotted), and a mixed galaxy sample of different galaxies (solid).
              }
         \label{Fig15}
   \end{figure}
%

   \begin{figure}
   \centering
   \includegraphics[angle=-90,width=88mm]{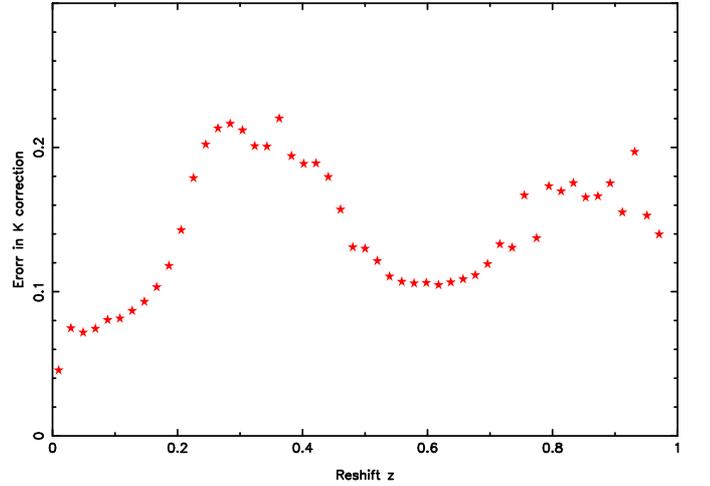}
      \caption{Average uncertainty in $K$ correction as a function of the redshift of galaxies.
        The uncertainty in $K$ correction is calculated from the uncertainty in redshift. The unit of error in $K$ correction is mag.
              }
         \label{Fig16}
   \end{figure}
%

   \begin{figure}
   \centering
   \includegraphics[angle=-90,width=88mm]{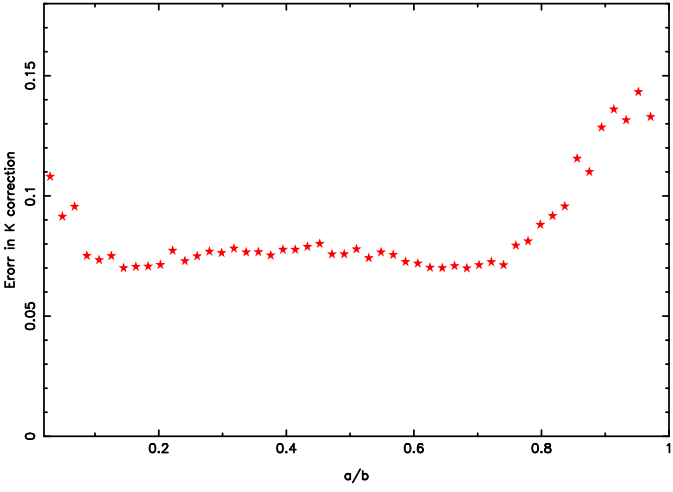}
      \caption{Similar to Fig. 16, but for the uncertainty in $K$ correction
        as a function of the ratio of semimajor to semiminor axis.
              }
         \label{Fig17}
   \end{figure}
%

   \begin{figure}
   \centering
   \includegraphics[angle=-90,width=88mm]{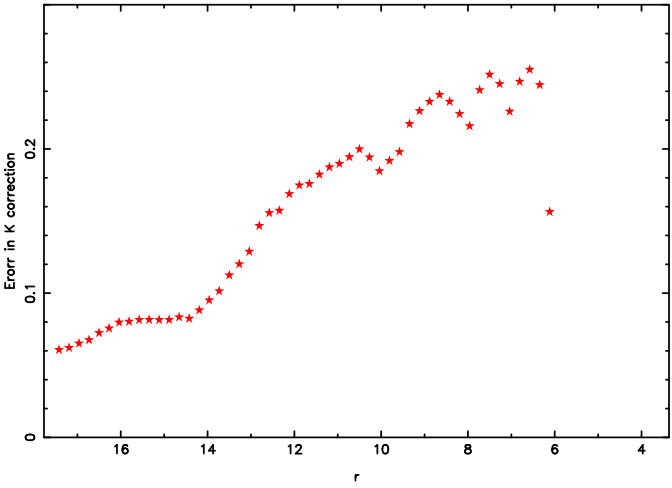}
      \caption{Similar to Fig. 16, but for the uncertainty in $K$ correction
        as a function of Petrosian magnitude in $r$ band. The unit of Petrosian magnitude is mag.
              }
         \label{Fig18}
   \end{figure}
%

   \begin{figure}
   \centering
   \includegraphics[angle=-90,width=88mm]{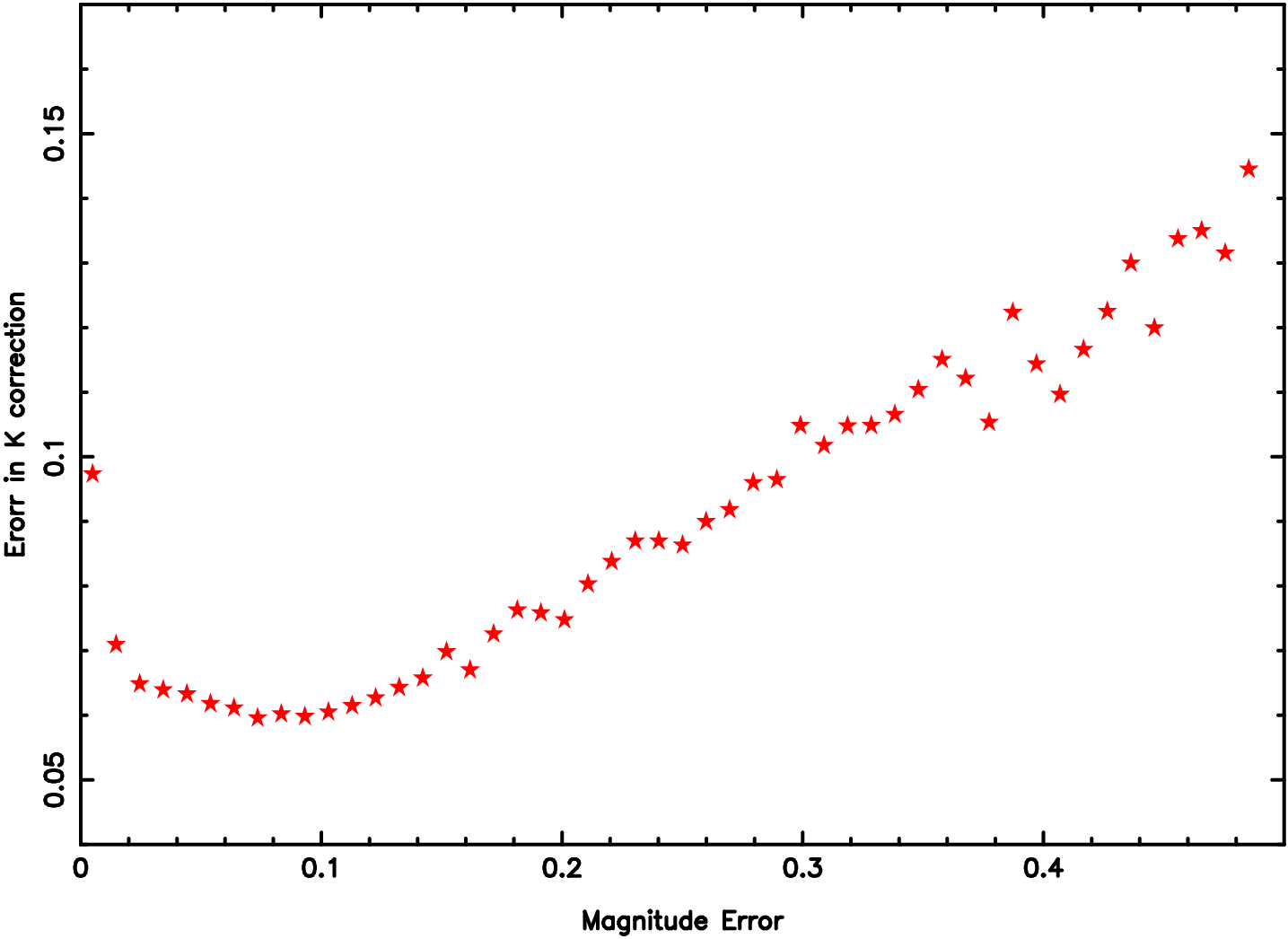}
      \caption{Similar to Fig. 16, but for the uncertainty in $K$ correction
        as a function of uncertainty of Petrosian magnitude in $r$ band.
              }
         \label{Fig19}
   \end{figure}
%

   In addition, in order to show whether two colours can determine the absorption line indices better or not, we compared the correlations between absorption line indices and different colours. The example results are shown in Figs. 20 and 21. In Fig. 20, the relations among Mg${\rm _b}$, $(u-z)$ and $(g-z)$ are plotted, while Fig. 21 is for the correlations among D$_{\rm n}$ (4000), $(u-r)$ and $(g-r)$. As can be seen, the two figures show clearly that, when taking two colours to estimate an age or metallicity sensitive absorption line index, the uncertainty does not become smaller, compared to the case using only one colour. Therefore, it is unnecessary to use two or more colours to estimate absorption line indices of galaxies.

   \begin{figure}
   \centering
   \includegraphics[angle=0,width=110mm]{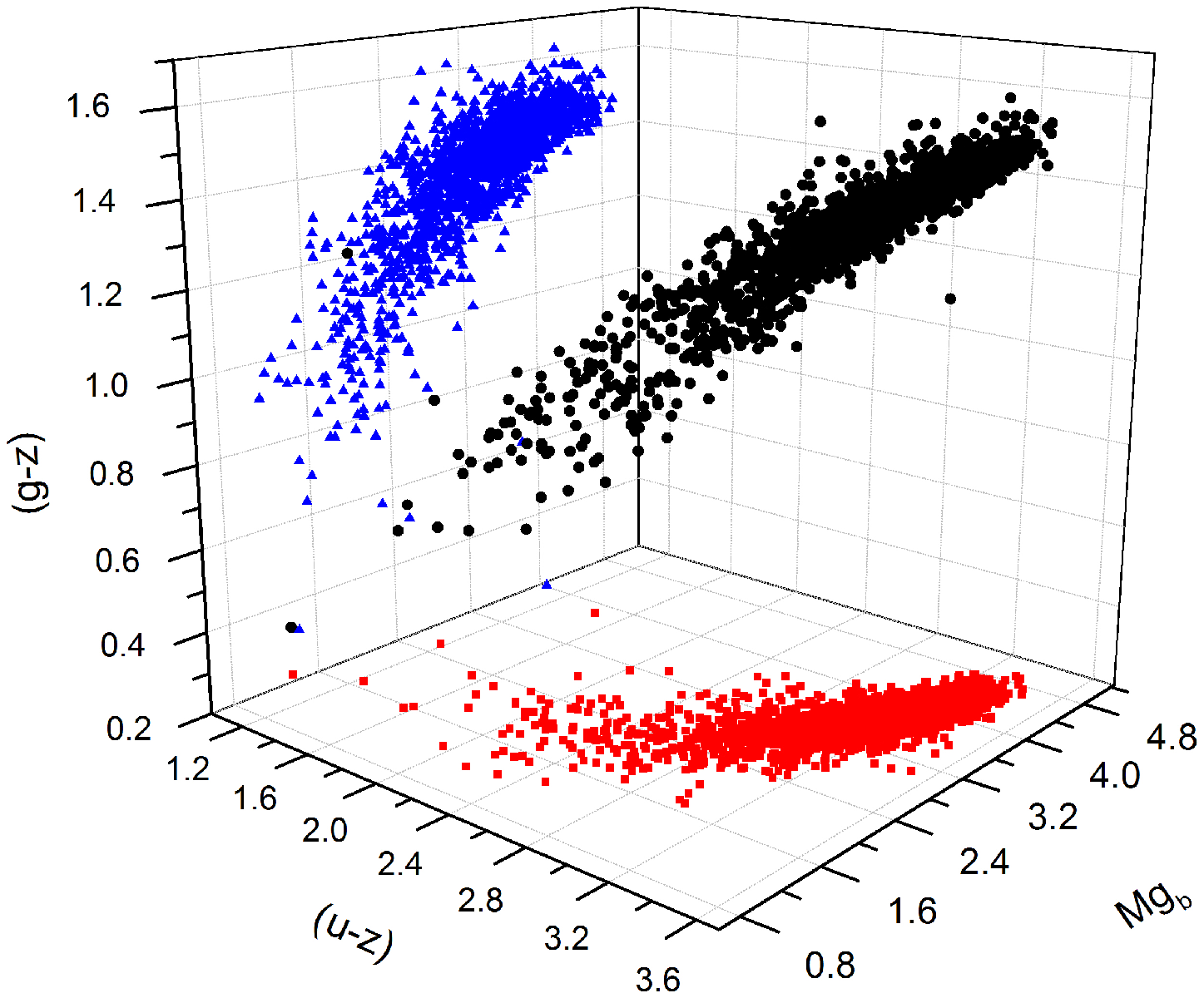}
      \caption{Relations among Mg$_{\rm b}$, $(u-z)$ and $(g-z)$. Colours and Mg$_{\rm b}$ index are in mag.
              }
         \label{Fig20}
   \end{figure}
%

   \begin{figure}
   \centering
   \includegraphics[angle=0,width=110mm]{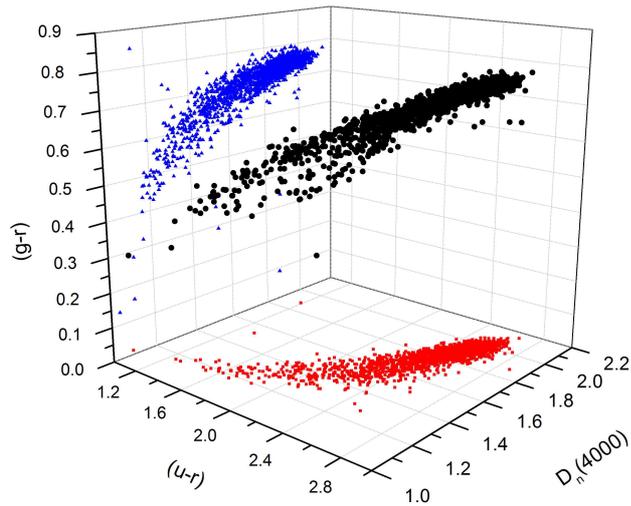}
      \caption{Similar to Fig. 20, but for D$_{\rm n}$(4000), $(u-r)$ and $(g-r)$. D$_{\rm n}$(4000) index is in $\AA$.
              }
         \label{Fig21}
   \end{figure}
%

\begin{table*}[]
\caption[]{Result uncertainties when using best-fit correlations to
estimate the absorption line indices of galaxy samples within
various redshift ranges. } \label{Tab2}
\begin{center}\begin{tabular}{llccc}
\hline \hline\noalign{\smallskip}
Correlation  & Redshift Range & Galaxy Number & $\sigma$ & Unit\\    
\hline
\\
Mg$_{\rm 1}$=~0.021+0.0004~e$^{(u-z)/0.612}$  & $z <$ 0.05 & 411 & 0.016 & mag\\
Mg$_{\rm 1}$=~0.021+0.0004~e$^{(u-z)/0.612}$  & 0.05 $\le z \le$ 0.10 & 1626 & 0.015 & mag\\
Mg$_{\rm 1}$=~0.021+0.0004~e$^{(u-z)/0.612}$  & $z >$ 0.10 & 250 & 0.016 & mag\\
Mg$_{\rm 2}$=~0.045+0.006~e$^{(u-z)/0.920}$   & $z <$ 0.05 & 411 & 0.027 & mag\\
Mg$_{\rm 2}$=~0.045+0.006~e$^{(u-z)/0.920}$   & 0.05 $\le z \le$ 0.10 & 1626 & 0.026 & mag\\
Mg$_{\rm 2}$=~0.045+0.006~e$^{(u-z)/0.920}$   & $z >$ 0.10 & 250 & 0.024 & mag\\
Mg$_{\rm b}$=~1.133+0.124~e$^{(u-z)/1.046}$   & $z <$ 0.05 & 411 & 0.435 & mag\\
Mg$_{\rm b}$=~1.133+0.124~e$^{(u-z)/1.046}$   & 0.05 $\le z \le$ 0.10 & 1626 & 0.394 & mag\\
Mg$_{\rm b}$=~1.133+0.124~e$^{(u-z)/1.046}$   & $z >$ 0.10 & 250 & 0.471 & mag\\
H$\delta_{\rm A}$=13.874-1.906~e$^{(g-r)/0.359}$  & $z <$ 0.05 & 411 & 1.697 & $\AA$\\
H$\delta_{\rm A}$=13.874-1.906~e$^{(g-r)/0.359}$  & 0.05 $\le z \le$ 0.10 & 1626 & 1.372 & $\AA$\\
H$\delta_{\rm A}$=13.874-1.906~e$^{(g-r)/0.359}$  & $z >$ 0.10 & 250 & 1.841 & $\AA$\\
H$\gamma_{\rm A}$=15.154-3.011~e$^{(g-r)/0.392}$    & $z <$ 0.05 & 411 & 1.855 & $\AA$\\
H$\gamma_{\rm A}$=15.154-3.011~e$^{(g-r)/0.392}$    & 0.05 $\le z \le$ 0.10 & 1626 & 1.518 & $\AA$\\
H$\gamma_{\rm A}$=15.154-3.011~e$^{(g-r)/0.392}$    & $z >$ 0.10 & 250 & 2.138 & $\AA$\\
H$\delta_{\rm F}$=3.266+16.141~$(g-r)$-26.131~${(g-r)}^2$   & $z <$ 0.05 & 411 & 0.698 & $\AA$\\
H$\delta_{\rm F}$=3.266+16.141~$(g-r)$-26.131~${(g-r)}^2$   & 0.05 $\le z \le$ 0.10 & 1626 & 0.713 & $\AA$\\
H$\delta_{\rm F}$=3.266+16.141~$(g-r)$-26.131~${(g-r)}^2$   & $z >$ 0.10 & 250 & 0.927 & $\AA$\\
H$\gamma_{\rm F}$=10.989-1.934~e$^{(g-r)/0.408}$    & $z <$ 0.05 & 411 & 1.158 & $\AA$\\
H$\gamma_{\rm F}$=10.989-1.934~e$^{(g-r)/0.408}$    & 0.05 $\le z \le$ 0.10 & 1626 & 0.903 & $\AA$\\
H$\gamma_{\rm F}$=10.989-1.934~e$^{(g-r)/0.408}$    & $z >$ 0.10 & 250 & 1.271 & $\AA$\\
D$_{\rm n}$(4000)=0.892+0.059~e$^{(u-r)/0.896}$    & $z <$ 0.05 & 411 & 0.102 & $\AA$\\
D$_{\rm n}$(4000)=0.892+0.059~e$^{(u-r)/0.896}$    & 0.05 $\le z \le$ 0.10 & 1626 & 0.103 & $\AA$\\
D$_{\rm n}$(4000)=0.892+0.059~e$^{(u-r)/0.896}$    & $z >$ 0.10 & 250 & 0.097 & $\AA$\\
\\
 \noalign{\smallskip}\hline
\end{tabular}\end{center}
\end{table*}

    The absorption line indices used in this paper rely on
    the STELIB spectral library (\citealt{LeBorgne:2003}). When other
    spectral libraries are used for defining absorption line indices,
    the differences caused by spectral libraries should be taken into account, while the effects of spectral libraries on colours can be neglected (see, e.g., \citealt{Bruzual:2003}).
    Therefore, before using the correlations found by this work to estimate absorption line
    indices, we suggest transforming the absorption line indices to
    the STELIB library \citep{Prugniel:2001}. A simple way is to use the offsets \citep{Johansson:2010} between indices on the Lick/IDS library (\citealt{Burstein:1984}; \citealt{Faber:1985}) and those on
    spectral libraries including MILES (\citealt{Sanchez-Blazquez:2006}), ELODIE (\citealt{Prugniel:2001}), and STELIB.
    However, it seems unnecessary to do transformations for some widely used indices such as H$\gamma_{\rm F}$, H$\beta$, Mg$_{\rm b}$, Fe5270, Fe5335, Fe5406, Fe5709, Fe5782, and Na$_{\rm D}$,
    because the STELIB, MILES, ELODIE, and Lick/IDS libraries give close values for these absorption line indices \citep{Johansson:2010}.

\section{Conclusions}

    We have investigated the correlations between absorption line
    indices and $ugriz$ colours using observational data relating to SDSS and a widely used stellar population model.
    Both observational data of early-type galaxies and a theoretical stellar population model show good correlations between
    some line indices and $ugriz$ colours of stellar populations.
    These correlations can be fitted as linear, exponential, or polynomial relations. The fitting
    uncertainties are found to be similar to, but somewhat larger than, the metrical uncertainties of line indices.

    In particular, it was found that some widely used metallicity and age-sensitive line
    indices,
    such as Mg$_{\rm b}$, Fe$_{\rm {5270}}$, Fe$_{\rm {5335}}$, H$\delta_{A}$,
    H$\gamma_{\rm A}$, H$\gamma_{\rm F}$, and D$_{\rm n}$(4000), are well correlated to some $ugriz$ colours. The
    relative uncertainties caused by the fitted correlations are small when transforming colours into absorption line
    indices.

    The results suggest that some estimates of absorption line indices can be obtained from the colours of galaxies.
    Our results are possibly useful for estimating the absorption
    line indices of galaxies that have no reliable spectral data
    available, and then for studying the stellar metallicities,
    ages, and element abundances.

\begin{acknowledgements}
We gratefully acknowledge Profs. R. Peletier and Gang Zhao for
constructive suggestions on this work. We also thank E. Casuso,
Prof. Zongwei Li, and Jinliang Hou for useful suggestions and the
MPA of Germany for sharing their data on the internet. This work has
been supported by Chinese Postdoctoral Science Foundation, Chinese
National Science Foundation (Grant No. 10963001), and the Yunnan
Science Foundation (Composite Population Studies of Galaxies Basing
on Photometric Data). Zhongmu Li  gratefully  acknowledges the
support of K. C. Wong  Education  Foundation, Hong Kong. Funding for
the SDSS and SDSS-II was provided by the Alfred P. Sloan Foundation,
the Participating Institutions, the National Science Foundation, the
U.S. Department of Energy, the National Aeronautics and Space
Administration, the Japanese Monbukagakusho, the Max Planck Society,
and the Higher Education Funding Council for England. The SDSS was
managed by the Astrophysical Research Consortium for the
Participating Institutions.
\end{acknowledgements}


\end{document}